\documentstyle[12pt,epsf]{article}
\newcommand{\sect}[1]{\setcounter{equation}{0}\section{#1}}

\def\la{\lambda}
\def\be{\begin{equation}}
\def\ee{\end{equation}}
\def\ba{\begin{eqnarray}}
\def\ea{\end{eqnarray}}
\def\bq{\begin{quote}}
\def\eq{\end{quote}}

\begin{document}
\thispagestyle{empty}
\begin{flushright}
Stanford SU-ITP-99-17
\\ April 1999
\end{flushright}
\vspace*{1cm}
\begin{center}
{\Large \bf Vacuum polarization in Schwarzschild space-time}\\
{\Large \bf by anomaly induced effective actions   }\\
\vspace*{2cm}
R. Balbinot, \footnote{E-mail: balbinot@bologna.infn.it}\\
\vspace*{0.2cm}
{\it Dipartimento di Fisica dell' Universit\`a di Bologna}\\
{\it and INFN sezione di Bologna}\\
{\it Via Irnerio 46, 40126 Bologna, Italy  }\\
\vspace*{0.5cm}
A. Fabbri, \footnote{E-mail:
afabbri1@leland.stanford.edu}\\
\vspace*{0.2cm}
{\it Department of Physics, Stanford University}\\
{\it Stanford, CA 94305-4060, USA}\\
\vspace*{0.5cm}
I. Shapiro \footnote{E-mail: shapiro@ibitipoca.fisica.ufjf.br }\\
\vspace*{0.2cm}
{\it Departamento de Fisica, ICE, Universidade Federal de Juiz de Fora}\\
{\it 36036-330, Juiz de Fora -- MG, Brazil } \\
\vspace*{0.2cm}
{\it Tomsk Pedagogical University, Tomsk, Russia}

\vspace*{2cm}
{ \bf Abstract}
\end{center}
The characteristic features of $\langle T_{\mu\nu}\rangle$ in the
Boulware, Unruh and Hartle-Hawking states for a conformal massless scalar
field propagating in the Schwarzschild space-time are obtained by means
of effective actions deduced by the trace anomaly. The actions are made
local by the introduction of auxiliary fields and boundary conditions are
carefully imposed on them in order to select the different quantum states.
 \vfill
\setcounter{page}{0}
\setcounter{footnote}{0}
\newpage
\sect{Introduction}
One of the most interesting aspects of quantum matter fields propagating
in curved space-time \cite{birdav}
 is the existence of a trace anomaly discovered by
Capper and Duff \cite{duff}
in 1974 and its deep connection with Hawking's black hole
evaporation \cite{Hawking}, \cite{duff-77}, \cite{chfu} . \\
At the classical level if the action describing the matter fields is
conformally invariant the trace $T$ of the corresponding stress energy
tensor vanishes. However, when the fields are quantized as a result of the
renormalization procedure the expectation value of the trace $\langle T
\rangle$ differs from zero when space-time dimension is even. We then have an
``anomaly''. This anomaly depends on geometrical quantities only: it is
expressed in terms of the Riemann tensor of the space-time and its
contractions. Furthemore being an ultraviolet effect, the anomaly is
general in the sense that its value does not depend on the quantum state
in which the expectation value is taken. In four dimensions (4D) we have
\be
\langle T^{\alpha}_{\alpha} \rangle \equiv \langle T \rangle =
-\frac{1}{(4\pi)^2}(aC^2 +bE +c\Box R )
\label{trac}
\ee
where $C^2\equiv C_{\alpha\beta\gamma\delta}C^{\alpha\beta\gamma\delta} $
is the square of the Weyl tensor
 and $E$ is an integrand of the Gauss-Bonnet topological term
\be
E\equiv R_{\mu\nu\alpha\beta}R^{\mu\nu\alpha\beta}
-4R_{\alpha\beta}R^{\alpha\beta} +R^2
\label{gabo}
\ee
and $a,b,c$ are known numerical coefficients that depend on the
spin of the matter fields considered \cite{birdav}. \\
In 2D we have the much simpler form
\be
\langle T^{\alpha}_{\alpha}\rangle \equiv \langle T\rangle = g R
\label{tradu}
\ee
where as before $g$ depends on the matter species. \\
The knowledge of the trace anomaly allows some information on the
complete quantum stress tensor expectation values $\langle
T_{\mu\nu}\rangle$ to be extracted. \\
In 2D the conservation equations $\nabla^a \langle T_a^{\ b} \rangle =0$
and the trace anomaly, eq. (\ref{tradu}), determine almost completely the
stress tensor.
Parametrizing the 2D spacetime metric $g_{ab}(x)$ as
\be
ds^2 =-e^{2\rho}dx^+dx^-
\label{tuco}
\ee
in the conformal gauge $\{ x^+, x^- \}$ one finds \cite{davies}
\ba
\langle x^{\pm} |T_{\pm\pm}|x^{\pm}\rangle &=& - \frac{1}{12\pi}
(\partial_{\pm}\rho\partial_{\pm}\rho - \partial_{\pm}^2\rho) +
t_{\pm}(x^{\pm}), \nonumber \\
\langle x^{\pm}|T_{+-}|x^{\pm}\rangle &=&
-\frac{1}{12\pi} \partial_+\partial_-\rho \ ,
\label{renst}
\ea
where $t_{\pm}(x^{\pm})$ are arbitrary functions of their arguments which
depend on the choice of quantum state in which the expectation values are
taken. They represent conserved radiation which, being traceless, cannot
be determined by the trace anomaly alone. These functions are fixed
by boundary conditions.
The above expressions for $\langle T_{ab}\rangle$ can also be obtained by
rigorous canonical quantization and point splitting regularization of the
divergences \cite{dafu}. \\
In 4D the situation is much more fuzzy since the four conservations
equations and the trace anomaly eq. (\ref{trac}) are clearly insufficient
to determine the ten components of $\langle T_{\mu\nu}\rangle$. Things
improve if the background space-time on which the quantum matter fields
propagate enjoys some symmetries which reduce the number of independent
components of $\langle T_{\mu\nu}\rangle$ \cite{chfu}. \\
Unfortunately in 4D it appears impossible (at least for the moment) to
find a nice formula analogous to eqs. (\ref{renst}) giving $\langle
T_{\mu\nu}\rangle $ as a function of a general metric $g_{\alpha\beta}$.
Such an expression would allow us the study of the evolution of the
background geometry driven by the quantum fluctuation of the matter
fields propagating on it. This is the so called backreaction, governed by
the semiclassical Einstein equations
\be
G_{\mu\nu} (g_{\alpha\beta})=8\pi \langle
T_{\mu\nu}(g_{\alpha\beta})\rangle \ , \label{i}
\ee
where $G_{\mu\nu}$ is the Einstein's tensor. \\
In principle $\langle T_{\mu\nu}(g_{\alpha\beta})\rangle$ can be derived
from an effective action $S_{eff}(g_{\alpha\beta})$ describing the
quantum matter fields propagating on $g_{\alpha\beta}$.
Although $S_{eff}(g_{\alpha\beta})$ is unknown, the trace anomaly
$\langle T\rangle$ can be used to reconstruct at least a part of it, i.e.
the so called ``anomaly induced effective action''.
$S_{eff}^{an}(g_{\alpha\beta})$ is then defined as following
\be
\frac{2}{\sqrt{-g}}g^{\mu\nu}\frac{\delta S_{eff}^{an}}{\delta g^{\mu\nu}}
=\langle T\rangle \ .
\label{llll}
\ee
By functional integration of this equation one obtains the explicit form
of $S_{eff}^{an}$. \\
The complete effective action $S_{eff}(g_{\alpha\beta})$ can then be
written as
\be
S_{eff}(g_{\alpha\beta})=S_{eff}^{an}(g_{\alpha\beta})+ S^W
(g_{\alpha\beta})\ ,
\label{coac}
\ee
where $S^W$ is a conformal invariant functional which can be regarded as
an integration constant of eq. (\ref{llll}). As expected, the
exact form of $S^W$ is not known. \\
Taking the functional derivation of $S_{eff}^{an}(g_{\alpha\beta})$ with
respect to $g_{\alpha\beta}$ we define the ``anomaly induced stress
tensor'' $S_{\mu\nu}$ as follows
\be
\frac{2}{\sqrt{-g}}\frac{\delta S_{eff}^{an}}{\delta g^{\mu\nu}}
=\langle S_{\mu\nu} \rangle
\label{stan}
\ee
where of course $\langle S\rangle = \langle T\rangle$.
It is tempting, given our ignorance on $\langle T_{\mu\nu}\rangle$,
to use $\langle S_{\mu\nu} \rangle$ in the semiclassical Einstein
equations (\ref{i}) to get at least a feeling on the backreaction
problem. \\
The basic question is: how close $\langle S_{\mu\nu}\rangle$ is to
$\langle T_{\mu\nu}\rangle $? In other words, can we ignore the Weyl
invariant part $S^W$ to get correctly at least the qualitative features
of the backreaction?
It is hard to find answers to the above questions since we do not know in
general $\langle T_{\mu\nu}(g_{\alpha\beta})\rangle$ . However, for some
particular background, say $\bar g_{\alpha\beta}$, $\langle T_{\mu\nu}
(\bar g_{\alpha\beta})\rangle$ is known and hence a check of the above
conjecture can be given by direct comparison of $\langle T_{\mu\nu}(\bar
g_{\alpha\beta})\rangle$ and the corresponding $\langle S_{\mu\nu} (\bar
g_{\alpha\beta})\rangle$. \\
The background $\bar g_{\alpha\beta}$ we have in mind is the Schwarzschild
black hole geometry
\be
ds^2=-(1-2M/r)dt^2 + (1-2M/r)^{-1}dr^2 +r^2 (d\theta^2 +\sin^2\theta
d\varphi^2)
\label{sci}
\ee
or
\be
ds^2 =-(1-2M/r) dudv + r^2 d\Omega^2\ ,
\label{edfi}
\ee
where
\ba
 u &=& t-r-2M\ln |r/2M -1|,
\nonumber \\
 v &=& t+r+2M\ln|r/2M -1|
\label{coed}
\ea
and $d\Omega^2$ is the metric for the unit two-sphere.
This, because of its physical implication, most notably Hawking black
hole evaporation, is one of the most studied space-times in quantum field
theory in curved space. \\
Detailed analytical and numerical investigations , by means of modes sum
and point-splitting regularization or by axiomatic principles,
of $\langle T_{\mu\nu}\rangle$ in this
background have been performed \cite{molti}.
We are rather confident on these results
on which, we stress, our understanding of all quantum black hole physics
is based. \\
In the Schwarzschild space-time three different quantum states
(``vacua'') are defined by expanding field operators in suitable basis:
\\ \\ \noindent
i) The Boulware vacuum $|B\rangle$ \cite{bula}. \\
This is obtained by choosing {\it in}
and {\it out} modes to be positive frequency with respect to the Killing
vector $\partial_t$ of the metric (\ref{sci}). This state most closely
reproduces the familiar notion of Minkowski vacuum $|M\rangle$
asymptotically. One finds, as $r \to \infty$, $\langle B|T_{\mu\nu}|B\rangle
\to 0 $ sufficiently rapidly, i.e. $O(r^{-6})$. Unfortunately,
the behaviour on
the horizon $r=2M$ is rather pathological, being $\langle
B|T_{\mu\nu}|B\rangle$ divergent there in a free falling frame. \\
 This
state is appropriate for the description of vacuum polarization around a
static star whose radius is necessary bigger than the horizon.
\\ \\\ \noindent
ii) The Unruh vacuum $|U\rangle$ \cite{uuru}. \\
 {\it in} modes are chosen as before to be
positive frequency with respect to $\partial_t$. This ensures that
asymptotically in the past $|U\rangle \sim |M\rangle$. On the other hand,
{\it out} modes are taken to be positive frequency with respect to Kruskal's
$U=-4Me^{-u/4M}$, the affine parameter along the past horizon. This mimics
the late time behaviour of modes coming out of a collapsing star as its
surface approaches the horizon. By this choice $\langle
U|T_{\mu\nu}|U\rangle$ is regular on the future (but not on the past) event
horizon and most remarkably asymptotically in the future $\langle
U|T_{\mu\nu}|U\rangle$ has the form of a flux of radiation at the Hawking
temperature $T_H=\frac{1}{8\pi M}$. \\ This state is the most appropriate to
discuss evaporation of black holes formed by gravitational collapse of
matter. In that case the divergence on the past horizon is spurious. \\ \\
\noindent iii) The Israel-Hartle-Hawking state $|H\rangle$ \cite{ishh}. \\
This state is obtained by choosing {\it in} modes to be positive frequency
with respect to Kruskal's $V=4Me^{v/4M}$, the affine parameter on the future
horizon, whereas outgoing modes are positive frequency with respect to $U$.
By this choice asymptotically both in the future and the past $|H\rangle \neq
|M\rangle$. \\ $\langle H|T_{\mu\nu}|H\rangle$ for $r\to\infty$ describes in
fact a thermal bath of radiation at the Hawking temperature $T_H$. By
construction this state is regular on both the future and the past horizon.
$|H\rangle$ is used to describe a black hole in thermal equilibrium with a
surrounding bath. \\ \\ We summarize here, for later comparison, the analytic
expression of the stress tensor for a conformally invariant scalar field in
the three vacua defined above. \\ In 2D Schwarzschild spacetime (neglect the
angular part in eq. (\ref{edfi}) ) we have \cite{full} \ba \langle B
|T_{tt}|B\rangle &=&  \frac{1}{12\pi} \left[ -\frac{2M}{r^3} +\frac{7 M^2}{2
r^4}\right] , \nonumber \\ \langle B |T_{rr}|B\rangle &=&
-\frac{1}{48\pi}\frac{2M^2}{r^4} (1-2M/r)^{-2}\ , \nonumber \\ \langle B
|T_{rt}|B\rangle &=& 0 \label{boli} \ea and \ba \langle U |T_{tt}|U\rangle
&=&  \frac{1}{12\pi} \left[ -\frac{2M}{r^3} +\frac{7M^2}{2r^4}\right] +
(768\pi M^2)^{-1} , \nonumber \\ \langle U |T_{rr}|U\rangle &=&
-\frac{1}{48\pi} (1-2M/r)^{-2}\frac{2M^2}{r^4} + (768\pi M^2)^{-1} ,
\nonumber \\ \langle U |T_{rt}|U\rangle &=& -(1-2M/r)^{-1}(768\pi M^2)^{-1}
\label{unli} \ea and finally \ba \langle H |T_{tt}|H\rangle &=&
\frac{1}{12\pi} \left[ -\frac{2M}{r^3} +\frac{7 M^2}{2 r^4}\right] + (384\pi
M^2)^{-1} , \nonumber \\ \langle H |T_{rr}|H\rangle &=&
-\frac{1}{48\pi}\frac{2M^2}{r^4} (1-2M/r)^{-2} + (384\pi M^2)^{-1} ,
\nonumber \\ \langle H |T_{rt}|H\rangle &=& 0 \ . \label{hhli} \ea Note that
$\langle B|T^a_a|B\rangle=\langle U|T^a_a|U\rangle= \langle H|T^a_a|H\rangle$
as a consequence of the state independence of the trace anomaly. \\ In 4D
only approximate expressions of $\langle T_{\mu\nu}\rangle$ can be given
\cite{molti}. Here we just report the leading behaviour at infinity and on
the horizon. For the Boulware vacuum $|B\rangle$ in Schwarzschild coordinates
$(t,r,\theta,\varphi)$
\be
\langle B| T_{\mu}^{\ \nu}|B\rangle \to O(r^{-6}) \ ,\ r\to\infty\ ,
\label{boin}
\ee
\be
\langle B| T_{\mu}^{\ \nu}|B\rangle \sim - \frac{1}{30\ 2^{12}\pi^2 M^4f^2}
\pmatrix{ -1 & 0 & 0 & 0
\cr 0 & 1/3 & 0 & 0
\cr 0 & 0 & 1/3 & 0
\cr 0 & 0 & 0 & 1/3 \cr}\ , \ r\to 2M\ ,
\label{boho}
\ee
where $f\equiv 1-2M/r$. \\
For the Unruh vacuum $|U\rangle$
\be
\langle U| T^{\mu}_{\nu}   |U\rangle \to \frac{L}{4\pi r^2}
\pmatrix{ -1 & -1 & 0 & 0
\cr 1 & 1 & 0 & 0
\cr 0 & 0 & 0 & 0
\cr 0 & 0 & 0 & 0 \cr},\ r\to\infty \ ,
\label {unin}
\ee
describing an outgoing flux, and
\be
\langle U| T_{a}^{\ b}   |U\rangle \sim  \frac{L}{4\pi (2M)^2}
\pmatrix{ 1/f & -1
\cr 1/f^2 & -1/f  \cr}, \ r\to 2M\ ,
\label {unho}
\ee
$a,b=r,t$,
which describes a negative energy flux of radiation going down the hole
compensating the escaping flux at infinity. $L$ is the luminosity. In
geometric optics approximation $L= \frac{2.197\ 10^{-4}}{\pi M^2} $
\cite{dewitt}.
Numerical extimates set $L=\frac{2.337\ 10^{-4}}{\pi M^2} $ \cite{elster}. \\
Finally for the $|H\rangle$ state
\be
\langle H| T^{\mu}_{\nu}   |H\rangle \to \frac{1}{30\ 2^{12}\pi^2 M^4}
\pmatrix{ -1 & 0 & 0 & 0
\cr 0 & 1/3 & 0 & 0
\cr 0 & 0 & 1/3 & 0
\cr 0 & 0 & 0 & 1/3 \cr}\ , \ r\to \infty\ ,
\label{hhin}
\ee
describing a thermal bath at the temperature $T_H$. Furthemore $\langle
H|T_{\mu}^{\ \nu}|H\rangle$ is regular on the horizons .
\\ \\ \noindent
In this paper, starting from the trace anomaly, we shall consider
$S_{eff}^{an}$ for a massless scalar field propagating in the
Schwarzschild space-time. From this, by functional derivation, we obtain
the corresponding stress tensor $\langle S_{\mu\nu}\rangle$ which should
then be compared with the expected results of $\langle T_{\mu\nu}\rangle$
given by eqs. (\ref{boli})-(\ref{hhin}).

\sect{Polyakov's action}

As a warm up exercise we start with the 2D Schwarzschild space-time and a
conformal massless scalar field propagating on it.
The purpose of it is purely pedagogical. The results will not be new,
although their nonstandard derivation is new and illustrates in a simple
way our strategy, the computational techniques and the basic physical
ingredients we shall use to deal the most interesting 4D case (see next
section). \\
The classical action for the massless field $f$ reads
\be
S=\int d^2x\sqrt{-g^{(2)}} \partial_{\mu} f\partial^{\mu}f\ ,
\label{scaca}
\ee
where $g^{(2)}$ is the determinant of the 2D Schwarzschild metric (see
eqs. (\ref{sci}) and (\ref{edfi}) with angular part omitted).
By canonical quantization and renormalization the corresponding $\langle
T_{ab}\rangle$ can be computed in the three vacua and the results are
given in eqs. (\ref{boli})-(\ref{hhli}). \\
Our procedure starts with the expression of the trace anomaly, which
in arbitrary 2D spacetime reads \cite{duff-77}
\be
\langle T\rangle = (24\pi)^{-1} R\ .
\label{antr}
\ee
The anomaly induced effective action is obtained by functional
integration of this anomaly (see eq. (\ref{llll}) ).
It is the well known Polyakov action \cite{polyak}
\be
 S_{eff}^{an}=S_P=-\frac{1}{96\pi } \int d^2 x \sqrt{-g} R \frac{1}{\Box}
R \ ,
\label{viii}
\ee
where $\Box$ is the d'Alembertian. \\
This action is nonlocal, but it can be made local either by choosing a
conformal gauge or by introducing an auxiliary field $\psi$ (and keeping
the gauge arbitrary). We can write
\be
S_P=-\frac{1}{96\pi} \int d^2x\sqrt{-g}\, (-\psi\Box\psi +2\psi R)\ .
\label{polo}
\ee
The equation of motion for $\psi$ is
\be
\Box\psi=R \ .
\label{empsi}
\ee
Once this is substituted back in eq. (\ref{polo}) we reobtain the
nonlocal form of eq. (\ref{viii}).
The anomaly induced stress tensor $\langle S_{ab}\rangle$ is defined as
\be
\frac {2}{\sqrt{-g}}\frac{\delta S_P}{\delta g^{ab}}\equiv \langle
S_{ab}\rangle = -\frac{1}{48\pi} \left[ 2\nabla_a\nabla_b\psi -
\nabla_a\psi \nabla_b\psi - g_{ab}\left(2R -\frac{1}{2}
(\nabla\psi)^2\right)\right] \ .
\label{stlo}
\ee
Our strategy is to solve eq. (\ref{empsi}) for the auxiliary field $\psi$
by assuming the 2D Schwarzschild spacetime as background, substitute the
result in eqs. (\ref{stlo}) to find $\langle S_{ab}\rangle$ and compare
with eqs. (\ref{boli})-(\ref{hhli}). \\
Before we start the program we have to solve a basic problem, how to
implement state dependence in our $\langle S_{ab}\rangle$. Being $\langle
T\rangle$ state independent, $S_{eff}^{an}$ makes no apparent reference
to a particular quantum state. \\
Note however that eq. (\ref{empsi}) determines the solution for $\psi$
modulo arbitrary solution of the homogeneous equation $\Box\psi=0$. It is
through the choice of the homogeneous solution that the state dependence
will be encoded in our framework. The homogeneous solution will be chosen
by imposing appropriate boundary conditions on $\psi$ that reflect the
physical features of the states $|B\rangle$, $|U\rangle$ and $|H\rangle$.\\
Since by eq. (\ref{empsi}) we have
\be
\psi= \frac{1}{\Box}R \ ,
\label{mmmm}
\ee
by this
procedure we impose boundary conditions on the operator $1/\Box$
characterizing the particular quantum state we are working with. \\
For the 2D Schwarzschild spacetime eq. (\ref{empsi}) becomes
\be
\partial_{a}[\partial^{a}\psi]=\frac{4M}{r^3}\ ,
\label{emca}
\ee
whose general solution we can write as
\be
\psi= at -\ln(1-2M/r) + A\left[ r+2M\ln(r-2M)\right] +B
\label{soge}
\ee
with
\be
\frac{\partial \psi}{\partial r} = (1-2M/r)^{-1} (-\frac{2M}{r^2} +A)\ ,
\label{depi}
\ee
where $a,A$ and $B$ are arbitrary constants which once fixed determine
the solution of the homogeneous equation. \\
The choice of a linear dependence in $t$ is simple. Given the structure
of $\langle S_{ab}\rangle$ in terms of $\psi$ (see eq. (\ref{stlo}) ), it
is clear that any $t$ dependence different from the linear would imply
$\partial_t \langle S_{ab}\rangle \neq 0$ , which contradicts the exact
result $\partial_t \langle T_{ab}\rangle =0$. The presence of such a
linear term allows, as we shall see, that $\langle S_{rt}\rangle \neq 0$
which is indeed needed in the Unruh state. \\
Inserting the solution for $\psi$ in eq. (\ref{stlo}) we obtain
\ba
\langle S_{tt}\rangle &=&  \frac{1}{12\pi}
\left[ -\frac{2M}{r^3} +\frac{7 M^2}{2 r^4}\right] + (48\pi)^{-1}
\frac{A^2 +a^2}{2}\ ,
\nonumber \\
\langle S_{rr}\rangle &=&
-\frac{1}{48\pi} (1-2M/r)^{-2}\left(\frac{2M^2}{r^4} -
\frac{A^2 +a^2}{2}\right)\ ,
\nonumber \\
\langle S_{rt} \rangle &=& (48\pi)^{-1}(1-2M/r)^{-1}Aa \ .
\label{solro}
\ea
Now we have to fix the arbitrary constants according to our choice of
quantum state.  \\ \\ \noindent
i) Boulware vacuum $|B\rangle$. \\
This state by construction reduces as
$r\to \infty$ to Minkowski vacuum and there $\langle B|T_{ab}|B\rangle$
should vanish. It is clear from the above equations that in order to
fullfill the asymptotic requirement we have to impose $a=A=0$ and we find
\ba
\langle B|S_{tt}|B\rangle &=&  \frac{1}{12\pi}
\left[ -\frac{2M}{r^3} +\frac{7M^2}{2r^4}\right] \ ,
\nonumber \\
\langle B| S_{rr}|B\rangle &=&
-\frac{1}{48\pi}\frac{2M^2}{r^4} (1-2M/r)^{-2} \ ,
\nonumber \\
\langle B| S_{rt}|B \rangle &=& 0 \ .
\label{bolio}
\ea
We see the exact agreement of our $\langle B|S_{ab}|B\rangle$ with eqs.
(\ref{boli}), namely
\be
\langle B|S_{ab}|B\rangle = \langle B|T_{ab}|B\rangle \ .
\label{exag}
\ee
With the above choice of integration constants the auxiliary field $\psi$
becomes
\be
\psi=-\ln(1-2M/r) +B
\label{sobo}
\ee
which vanishes for $r\to\infty$ if we further set $B=0$. Note the
singularity of $\psi $ for $r=2M$  which causes the divergence of
$\langle B|S_{ab}|B\rangle $ on the horizon. \\ \\ \noindent
ii) Unruh state $|U\rangle$. \\
 By assumption there is no incoming flux
(i.e. $\langle U|S_{vv}|U\rangle=0$ as $r\to\infty$) and $\langle
U|S_{ab}|U\rangle$ has to be regular on the future horizon (i.e.
$\langle U|S_{uu}|U\rangle \to 0$ at least as $(r-2M)^2$, $\langle
U|S_{vv}|U\rangle \sim reg.$ and $\langle U|S_{uv}|U\rangle \sim (r-2M)$
for $r\to 2M$ ).
The asymptotic condition
\be
\langle U|S_{vv}|U\rangle = \frac{1}{4}\left( \langle U|S_{tt}|U\rangle
+ (1-2M/r)^2\langle U|S_{rr}|U\rangle +2(1-2M/r)\langle
U|S_{rt}|U\rangle \right) \to 0
\label{noin}
\ee
requires $A=-a$. Furthemore $\langle U|S_{uu}|U\rangle$ vanishes like
$(r-2M)^2$ for $r\to 2M$ if $A=(4M)^{-1}$.
Summarizing we have
\ba
\langle U |S_{tt}|U\rangle &=&  \frac{1}{12\pi}
\left[ -\frac{2M}{r^3} +\frac{7M^2}{2r^4}\right] + (768\pi M^2)^{-1} ,
\nonumber \\
\langle U |S_{rr}|U\rangle &=&
- (1-2M/r)^{-2}\left[ \frac{M^2}{24\pi r^4} - (768\pi M^2)^{-1}\right] ,
\nonumber \\
\langle U |S_{rt}|U\rangle &=& -(1-2M/r)^{-1}(768\pi M^2)^{-1} \ .
\label{anin}
\ea
Comparison with eqs. (\ref{unli}) reveals
\be
\langle U|S_{ab}|U\rangle = \langle U|T_{ab}|U\rangle\ .
\label{coun}
\ee
It is quite interesting to examine the behaviour of the auxiliary field
$\psi$. From eq. (\ref{soge}) we see that the asymptotic requirement
$A=-a$ yields $\psi \sim -Au $ as $r\to \infty$, whereas the regularity
condition on the horizon $A=(4M)^{-1}$ reveals that $\psi \sim -v/4M +
const.$ as $r\to 2M$.
We have therefore the nice connection: \\ \noindent
a) for $r\to\infty$ (we set $B=0$) $\psi \sim -u/4M \
\Leftrightarrow
\langle U |S_{ab}|U\rangle $ describes outgoing radiation;
\\ \noindent
b) for $r\to 2M$ $\psi \sim -v/4M +const. \ \Leftrightarrow
\langle U |S_{ab}|U\rangle$ describes an ingoing flux of negative
energy radiation; $\psi$ and $\partial_a\psi$, $a,b=v,r$ regular on the
future horizon $\Leftrightarrow \ \langle U |S_{ab}|U\rangle$ regular on the
future horizon. \\ \\
iii) Israel-Hartle-Hawking state $|H\rangle$. \\
This state corresponds to
thermal equilibrium and thus we demand $\psi$ to be time independent,
i.e. $a=0$ which implies $\langle H|S_{rt}|H\rangle=0$. Furthemore
$|H\rangle$ is constructed in a way that makes the stress tensor regular
both on the future and past horizon, which is achieved by requiring that
as $r\to 2M$ $\langle H|S_{uu}|H\rangle=\langle H|S_{vv}|H\rangle$ vanish
like $(r-2M)^2$. Inspection of eqs. (\ref{solro}) reveals that we can
fullfill the above requirement by demanding
\be
\frac{1}{48\pi} \frac{A^2}{2}  = \frac{1}{384\pi M^2} \ ,
\label{prul}
\ee
i.e.
\be
A=\frac{1}{2M}\ .
\label{afiss}
\ee
With the above choice $\psi$ is regular as $r \to 2M$, since by eq.
(\ref{soge}) we have that for the state $|H\rangle$
\be
\psi= \ln r + \frac{r}{2M} +B
\label{psihh}
\ee
which is indeed regular.
\footnote{The other solution of eq. (\ref{prul}), namely
$A=-1/2M$, makes $\psi$ singular on the horizon.}
We can make $\psi$ vanishing on the horizon by choosing the arbitrary
constant $B$ such that
\be
B=-(\ln 2M +1)\ ,
\label{bico}
\ee
i.e. for the $|H\rangle$ state we have
\be
\psi= \ln r +\frac{r}{2M} - (\ln2M +1)\ .
\label{pire}
\ee
Note that this expression could have been derived by looking at the static
solution of the auxiliary field equation (\ref{emca}) of the form
\be
\psi(r)= \int_{2M}^r dr_1 (1-2M/r_1)^{-1}\int _{2M}^{r_1} dz
\frac{4M}{z^3} \ .
\label{conint}
\ee
Inserting now $a=0$ and $A=1/2M$ in eqs. (\ref{solro}) we find
\ba
\langle H|S_{tt}|H\rangle &=& (12\pi)^{-1}\left[-\frac{2M}{r^3}
+\frac{7M^2}{2r^4}\right] + (384\pi M^2)^{-1}\ ,
\nonumber \\
\langle H|S_{rr}|H\rangle &=& -(48\pi)^{-1} (1-2M/r)^2 \left[
\frac{2M}{r^4} - \frac{1}{8M^2} \right]\, \nonumber \\
\langle H|S_{tr}|H\rangle &=& 0 \ .
\label{ricu}
\ea
Direct check with eqs. (\ref{hhli}) reveals again the identity
\be
\langle H|S_{ab}|H\rangle= \langle H|T_{ab}|H\rangle\ .
\label{ulid}
\ee
We have therefore shown in this simple 2D example the perfect agreement
of our $\langle S_{ab}\rangle$ and $\langle T_{ab}\rangle$. This is not
surprising at all; it is a general feature not at all limited to the 2D
Schwarzschild background. It is well known that the auxiliary field
equation can be integrated without making reference to any particular
background.
An arbitrary 2D spacetime
\be
ds^2 =-g_{ab}dx^adx^b
\label{geme}
\ee
can always be parametrized as follows (conformal gauge)
\be
ds^2=-e^{2\rho}dx^+dx^- \ .
\label{conga}\ee
In that gauge the general solution of the equation for $\psi$ (eq.
(\ref{empsi}) ) reads
\be
\psi= -2\rho + F(x^+) + G(x^-)\ ,
\label{roso}
\ee
where $F(x^+)$ and $G(x^-)$ are arbitrary functions of their arguments.
Inserting this in eq. (\ref{stlo}) we obtain $\langle S_{ab}\rangle$.
This has exactly the same form as $\langle T_{ab}\rangle$ in eqs.
(\ref{renst}) which, as already said, can be derived by canonical
quantization. \\
The motivation for our nonstandard derivation of the stress tensor for
a conformal scalar in 2D Schwarzschild space-time was purely
pedagogical\footnote{More detailed consideration, including
critical analysis of the current literature, have been reported
separately in \cite{NEW}.}.
It has shown explicitly how to implement state dependence in the
framework of the anomaly induced effective action by imposing appropriate
boundary conditions on the auxiliary field $\psi$ (or equivalently on
$\langle S_{ab}\rangle$) which reflect the physical properties of the
quantum state.

\sect{Four dimensions: Reigert's action}

Having set the basis of our formalism we are ready to attack the
physically more interesting example: a conformal invariant scalar field
$f$ in the 4D Schwarzschild black hole space-time.
The classical action describing massless conformal invariant
scalar field $f$ is
\be
S=\int d^4x\sqrt{-g}\left( \partial_{\mu}f \partial^{\mu}f
-\frac{1}{6} Rf^2\right) + S_{vacuum}\, ,
\label{quca}
\ee
where the $S_{vacuum}$ is necessary for the renormalizability of the
theory  (see \cite{book} for the general introduction to the
renormalization in curved space-time).
Performing the path integration over the scalar field, one
meets divergences and the renormalization is necessary.
The trace anomaly results from the renormalization
of the vacuum action in (\ref{quca}).

The vanishing of the classical trace has its quantum counterpart
in the presence of a trace anomaly, composed by the terms
coming from the vacuum counterterms \cite{duff-77}
\begin{equation}
\label{6}
\langle T^{\alpha}_{\alpha}\rangle \equiv \langle T \rangle =-
\frac{1}{(4\pi)^2}\left( aC^2 + bE +c\Box R \right) \ .
\end{equation}

For the free conformal invariant matter fields the three structures presented
in (\ref{6}) are sufficient, and the $\,\sqrt{-g}R^2$-type counterterm does
not arise. That is why the corresponding term is absent in (\ref{6})
\footnote{Also we notice that that there is no expression, local or nonlocal,
in the effective action, which could produce the term $\,\sqrt{-g}R^2$ in the
trace anomaly \cite{reigert}.}.

The anomaly induced effective action which yields the anomaly
(\ref{6})
has been independently derived by Reigert \cite{reigert}
and Fradkin and Tseytlin \cite{frts}
(see also \cite{book,antmot,cosh,anju} for further
applications and references about the Reigert's action).
The expression derived in \cite{reigert} has the form:
\ba
S_{eff}^{an}=
\frac{1}{(4\pi)^2} \int d^4 x \int d^4 y \left\{\sqrt{-g}
\left[ E-\frac{2}{3}\Box R\right]\right\}_x G(x,y)
\nonumber \\
 \left\{ \sqrt{-g}\left[ \frac{a}{4}C^2
+\frac{b}{8}(E-\frac{2}{3}\Box R)\right]\right\}_y
- \frac{c+\frac{2}{3}b}{12 (4\pi)^2} \int d^4 x\sqrt{-g(x)}R^2\ ,
\label{reiac}
\ea
where $a= 1/120$, $b=-1/360$, $c = 1/180$.
$G(x,y)$ is the inverse of the fourth order
operator $\Delta_4$
\begin{equation}
\label{10}
\Delta_4=\Box^2 - 2R^{\mu\nu}\nabla_{\mu}\nabla_{\nu}+
\frac{2}{3}R\Box  -\frac{1}{3}(\nabla^{\mu}R)\nabla_{\mu} \ ,
\end{equation}
i.e.
\be
(\sqrt{-g}\Delta_4)_x G(x,y)=\delta^4(x-y) \ .
\label{11}
\ee
Note that $\Delta_4$ acting on a scalar field of zero conformal weight
is the unique self-adjoint
conformal invariant differential operator in $4D$ \cite{pan}.
This also implies the conformal invariance of $G(x,y)$.
Let us rewrite the Reigert's action in a more symmetric way
\ba
S_{eff}^{an} &=&- \frac{c+\frac{2}{3}b}{12(4\pi)^2}\int d^4 x \sqrt{-g}R^2
-\frac{1}{2(4\pi)^2}\int d^4 x \sqrt{-g(x)}
\int  d^4 y \sqrt{-g(y)}
\nonumber \\
&\ &\frac{\sqrt{-b}}{2} \left[ (E-\frac{2}{3}\Box
R)+\frac{a}{b}C^2\right]_x
G(x,y) \frac{\sqrt{-b}}{2}\left[ (E-\frac{2}{3}\Box R) +
\frac{a}{b}C^2\right]_y \nonumber \\
&+& \frac{1}{2(4\pi)^2}\int d^4 x\sqrt{-g(x)} \int d^4 y  \sqrt{-g(y)}
\left( \frac{a}{2\sqrt{-b}}C^2\right)_x
G(x,y)\left(\frac{a}{2\sqrt{-b}}C^2\right)_y \ .
\label{azsim}
\ea
This nonlocal action can be made local by the introduction of two
auxiliary fields $\phi$ and $\psi$ \cite{reigert,shapiro}
\ba
\label{14}
S_{eff}^{an}&=&
+\int d^4 x\sqrt{-g}\left[ \frac{1}{2}\phi\Delta_4\phi +
\phi\left(
\frac{\sqrt{-b}}{8\pi}(E-\frac{2}{3}\Box R)-\frac{a}{8\pi\sqrt{-b}}C^2
\right) \right] \nonumber \\
&+&\int d^4x\sqrt{g}\left(-\frac{1}{2}\psi\Delta_4\psi
+\frac{a}{8\pi\sqrt{-b}} C^2 \psi \right)
-\frac{c+\frac{2}{3}b}{12(4\pi)^2}\int d^4x \sqrt{-g}R^2\ .
\ea
Once the equations of motion for $\phi$ and $\psi$ are used, eq.
(\ref{14}) reduces to the nonlocal form (\ref{reiac}).
We remark that
the path integration over the auxiliary fields would
give an extra contribution to the anomaly (\ref{6}), modifying
the values of the constants $a,b,c$ in (\ref{6}). Here we
consider the introduction of the auxiliary fields $\psi,\phi$
as a method of working with the nonlocalities of the expression
(\ref{azsim}) and hence require the correspondence to hold
only on classical level.
\\
Note that the last nonlocal term in eq. (\ref{azsim}) (or equivalently the
second integrand in eq. (\ref{14}) ) is conformally invariant and hence could
be removed and included in the Weyl invariant part of the effective action
$S^W$ (see eq. (\ref{coac}) ). For later use we shall keep this term. However
its coefficient $a/(8\pi\sqrt{-b})$ can be replaced by an arbitrary
parameter, say $l_1$, since a change in this parameter can be reabsorbed by
the Weyl invariant part $S^W(g_{\alpha\beta})$ of $S_{eff}$ which is not
determined by the trace anomaly. \\ So we shall work with the following
anomaly induced effective action \ba \label{elle}
S_{eff}^{an}&=&-\frac{k_3}{12}\int d^4 x\sqrt{-g} R^2 +\int d^4
x\sqrt{-g}\left[ \frac{1}{2}\phi\Delta_4\phi + \phi\left(k_1C^2 +
k_2(E-\frac{2}{3}\Box R) \right) \right] \nonumber \\ &+&\int
d^4x\sqrt{g}\left(-\frac{1}{2}\psi\Delta_4\psi +l_1 C^2 \psi \right) \ , \ea
where we set $k_1=-a/(8\pi\sqrt{-b})$, $k_2=\sqrt{-b}/8\pi$ and
$k_3=(c+\frac{2b}{3})/(4\pi)^2 $, bearing in mind the values of $a,b,c$
specific for a single conformal scalar. \\ After eliminating the auxiliary
fields $\phi$ and $\psi$ eq. (\ref{elle}) does not reduce to Reigert's
nonlocal action of eq. (\ref{reiac}) (unless $l_1=a/(8\pi\sqrt{-b})$). The
difference, as we just remarked, is a conformally invariant functional, hence
of the form $S^W$, on which we have no control. \\ From eq. (\ref{elle}) the
equations of motion for the auxiliary fields follow
\begin{equation}
\label{15}
\frac{1}{\sqrt{-g}}\frac{\delta S_{eff}^{an}}{\delta\phi}=
\Delta_4\phi + k_1 C^2 + k_2(E-\frac{2}{3}\Box R)=0\ ,
\end{equation}
\begin{equation}
\label{16}
\frac{1}{\sqrt{-g}}\frac{\delta S_{eff}^{an}}{\delta\psi}=
 -\Delta_4\psi +l_1 C^2 =0\ .
\end{equation}
Introducing now the traceless tensor $K_{\mu\nu}(\phi)$
\ba
\label{17}
&\ & K_{\mu\nu}(\phi) = \frac{1}{\sqrt{-g}}\,\frac{\delta }{\delta
g^{\mu\nu}}
\int d^4x\sqrt{-g}\,\left\{ \phi\Delta_4\phi \right\}
= -\frac12\,g_{\mu\nu}\,\left(
(\Box \phi)^2 - 2R^{\alpha\beta}\phi_\alpha\phi_\beta +
\frac23\,(\nabla\phi)^2 R\right)  \nonumber \\
&-& \phi_\mu(\nabla_\nu {\Box}\phi)
- \phi_\nu(\nabla_\mu {\Box}\phi)
+ 2\phi_{\mu\nu}({\Box}\phi)
+ 2\phi^\lambda\phi_{\mu\nu\lambda}
- \frac23\,\phi^\lambda \,(\phi_{\lambda\mu\nu} + \phi_{\lambda\nu\mu})
+ \frac13\,g_{\mu\nu}\,\phi^{\lambda\tau}\,\phi_{\lambda\tau}
\nonumber \\
&+& \frac13\,g_{\mu\nu}\,\phi^{\la}\,({\Box}\phi_{\la})
- \frac43\,\phi^{\lambda}_\mu\,\phi_{\lambda\nu}
- 2\,\phi_\lambda\,(R^\lambda_\mu\phi_\nu + R^\lambda_\nu\phi_\mu)
+ \frac23\,R\,\phi_\mu\,\phi_\nu
+ \frac23\,R_{\mu\nu} \,\phi_\lambda\,\phi^\lambda
\ea
we can write the anomaly induced stress tensor $\langle
S_{\mu\nu}\rangle$ as
\ba
\label{18}
\frac{2}{\sqrt{-g}}\frac{\delta S^{eff}_{an}}{\delta g^{\mu\nu}}
&=& \langle S_{\mu\nu}\rangle = K_{\mu\nu}(\phi)-K_{\mu\nu}(\psi)
+ 8\nabla^{\lambda}\nabla^{\tau}Z R_{\mu\lambda\nu\tau} -\,g_{\mu\nu}\, Z
R^2_{\rho\sigma\alpha\beta}
\nonumber \\
&+& 4Z\,R_{\mu\rho\lambda\tau}\,{R_{\nu }}^{\rho\lambda\tau}
+\frac{4k_2}{3}\,\left[
(\nabla_\mu\nabla_\nu {\Box} \phi)
- g_{\mu\nu} ({\Box}^2 \phi) \right] +...
\ea
where we have defined
\begin{equation}
\label{19}
Z\equiv (k_1+k_2)\phi +l_1\psi
\end{equation}
and $\phi_{\lambda\tau}\equiv \nabla_{\lambda}\nabla_{\tau}\phi$, etc. .
The dots in eq. (\ref{18}) indicate terms containing either the Ricci
tensor $R_{\mu\nu}$ or its contraction $R$.
In our subsequent analysis the background space-time chosen will be
Schwarzschild  and therefore these terms, together with the sum of the two
terms
linear in $Z$ in eq. (\ref{18}), vanish identically. \\
The equation of motion for $\phi$ in the Schwarzschild spacetime (see eq.
(\ref{sci})) reads
\be
\Box_s^2\phi=\alpha \frac{M^2}{r^6}\ ,
\label{emsb}
\ee
where $\alpha \equiv -48(k_1+k_2)$ and $\Box_s$ is the Delambertian written in
Schwarzschild coordinates.
The solution for $\phi$ can be given in the following form
\be
\phi(r,t)=d\cdot t + w(r)\ ,
\label{capo}
\ee
where $w(r)$ is such that
\ba
\label{22}
\frac{dw}{dr} &=& \frac{B}{3}r +\frac{2}{3}MB-\frac{A}{6}-\frac{\alpha}{72M}
+ \left(\frac{4}{3}BM^2 + \frac{C}{2M}-AM -\frac{\alpha}{24}\right)
\frac{1}{r-2M}
-\frac{C}{2M}\frac{1}{r} \nonumber \\
&+&\ln r \left[ -\frac{\alpha M}{18}\frac{1}{r(r-2M)}
-\left(\frac{A}{2M}-\frac{\alpha}{48M^2}\right)\frac{r^2}{3(r-2M)} \right]
\nonumber \\
&+& \ln(r-2M) \left[ \left(\frac{A}{2M}-\frac{\alpha}{48M^2}\right)
\frac{r^3 -(2M)^3}{3r(r-2M)}
\right]
\ea
and $(d,A,B,C)$ are constants that specify the homogeneous solution
$\Box_s^2\phi=0$ and hence the quantum state. The presence of a linear
term in time in eq. (\ref{capo}) has the same explanation given in
section II (see the discussion following eq. (\ref{depi}); it allows
$\langle S_{rt}\rangle \neq 0$ but keeping $\partial_t \langle
S_{\mu\nu}\rangle=0$).
One can give solution for the equation of motion for the other auxiliary
field $\psi$ in the same form with obvious replacements
$$
\alpha \equiv - 48 (k_1+k_2) \longrightarrow \beta\equiv 48l_1
\,\,\,\,\,\,\,\,\, {\rm and}
\ (d,A,B,C)\longrightarrow (d',A',B',C')\,.
$$
Substituting the solutions for the auxiliary fields in eqs. (\ref{18}) we
obtain the effective stress tensor for the Schwarzschild space-time in
the form
\be
\langle S_{\mu\nu}\rangle = \langle S_{\mu\nu}(\phi)\rangle +
\langle S_{\mu\nu}(\psi)\rangle
\label{seva}
\ee
separating the contribution of each individual field to $\langle
S_{\mu\nu}\rangle$. Both $\phi$ and $\psi$ are related to the
inverse of
the fourth order differential operator $\Delta_4$, but they are
independently defined, and therefore the
boundary conditions
have to be imposed on them individually.
For later use
we reproduce the limiting behaviour of $\partial_r\phi$ as
\begin{equation}
\label{25}
\partial_r\phi \sim \frac{E}{r-2M} +2M
(\frac{A}{2M}-\frac{\alpha}{48M^2})\ln(r-2M) + reg.\ terms \ ,
\end{equation}
where
\begin{equation}
\label{26}
E=-\frac{\alpha}{12}+\frac{4}{3}BM^2 +\frac{C}{2M}-2M^2
(\frac{A}{2M}-\frac{\alpha}{48M^2})-\frac{2}{3}AM \ln2M \ ,
\end{equation}
in the limit $r\to 2M$, and
\begin{equation}
\label{29}
\partial_r\phi \to \frac{B}{3}r + \frac{2}{3}MB -\frac{A}{2}
\end{equation}
at infinity. \\ \\ \noindent
i) Boulware state $|B\rangle$.  \\ \noindent
For flat space and Minkowski vacuum $\phi=const.$ (which we can set to
zero without any consequences). This implies $\langle
M|S_{\mu\nu}|M\rangle=0$ as it should be.
Since as $r\to \infty$ $|B\rangle \to |M\rangle$, the correct asymptotic
limit on $\phi$ is obtained in eqs. (\ref{capo}), (\ref{22}) by setting
all constants to zero. So for the state $|B\rangle$ we have
\ba
\label{ccoo}
\frac{dw}{dr}&=& -\frac{\alpha}{72M}
-\frac{\alpha}{24}\frac{1}{r-2M}
+\ln r \left[ -\frac{\alpha M}{18}\frac{1}{r(r-2M)}
+\frac{\alpha}{48M^2}\frac{r^2}{3(r-2M)} \right] \nonumber \\
&+&\ln(r-2M) \left[ -\frac{\alpha}{48M^2}\frac{r^3 -(2M)^3}{3r(r-2M)}
\right]
\ea
and an analogous expression for $\psi$ ($\alpha\to \beta$). \\
Starting from these expressions one can obtain $\langle
B|S_{\mu\nu}|B\rangle$. Being the calculations rather long and boring, we
report just the limiting behaviours
\begin{equation}
\label{ciuco}
\langle B| S_{\mu}^{\nu}  |B \rangle \to  \frac{1}{2}
\frac{\alpha^2-\beta^2}{(24)^2}\frac{1}{(2M)^4f^2}
\pmatrix{ -1 & 0 & 0 & 0
\cr 0 & 1/3 & 0 & 0
\cr 0 & 0 & 1/3 & 0
\cr 0 & 0 & 0 & 1/3 \cr}
\end{equation}
for $r\to 2M$ ($f=1-2M/r$) and
\be
\langle B| S_{\mu}^{\nu}  |B \rangle \to O(\frac{1}{r^6})
\label{calo}
\ee
as $r\to \infty$. \\
The qualitative agreement of our $\langle B|S_{\mu\nu}|B\rangle$ and
 $\langle B|T_{\mu\nu}|B\rangle$ of eqs. (\ref{boin}), (\ref{boho}) is
rather nice. We see the expected $r^{-6}$ fall off and the $1/f^2$
divergence on the horizon. \\ \\ \noindent
ii) Unruh state $|U\rangle$. \\ \noindent
The Unruh state $|U\rangle$ agrees with Minkowski vacum on past null
infinity, i.e. no incoming radiation. This requires $\phi\sim u$ as
$r\to\infty$, which implies $\langle U|S_{vv}|U\rangle \to 0$ for $r\to
\infty$. This asymptotic behaviour of $\phi$ is achieved by requiring
$B=0$ and $d=A/2$. Regularity along the future horizon requires $\phi
\sim v$ and $\partial_a\phi\sim reg.$, $a=v,r$ (see the discussion
following eq. (\ref{coun})). This is achieved by fixing $E=2dM$ ($\phi\sim
v$) and $A=\alpha/24M$ ($\partial_a\phi$ finite). That this is
the correct
choice it can be seen by direct evaluation of $\langle
U|S_{\mu\nu}|U\rangle$ \cite{balfabsha}. \\
Near the horizon to the order $f^{-2}$ we have
\begin{equation}
\label{27}
\langle U| S_{\mu}^{\ \nu}(\phi)|U\rangle \sim
\frac{(E^2-4d^2M^2)}{32M^4f^2}
\pmatrix{ -1 & 0 & 0 & 0
\cr 0 & 1/3 & 0 & 0
\cr 0 & 0 & 1/3 & 0
\cr 0 & 0 & 0 & 1/3 \cr}\ ,
\end{equation}
which indeed vanishes for $E=2dM$. Logarithmic divergence in the pressure
$\langle T_{\theta}^{\ \theta}\rangle$ are eliminated by $A=\alpha/24M$.
Repeating the same arguments for $\psi$ gives $E'=2d'M, A'=\beta /24M, B'=0$.
The following leading behaviour of $\langle S_{ab}\rangle$ emerges
\begin{equation}
\label{28}
\langle U|  S_{a}^{\ b} |U \rangle \sim
\frac{\alpha^2-\beta^2}{2(48M^2)^2} \pmatrix{ 1/f & -1
\cr 1/f^2 & -1/f \cr}\ ,\ r\to 2M\ ,
\end{equation}
which is indeed regular on the future horizon. By the above choice of the
constants, we can find the asymptotic form of $\langle
U|S_{\mu\nu}|U\rangle$
\begin{equation}
\label{giango}
\langle U|  S_{\mu}^{\ \nu} |U  \rangle \to
\frac{\alpha^2-\beta^2}{2r^2(24M)^2}
\pmatrix{ -1 & -1 & 0 & 0
\cr 1 & 1 & 0 & 0
\cr 0 & 0 & 0 & 0
\cr 0 & 0 & 0 & 0
\cr}\ , \ r\to \infty \ .
\end{equation}
Our results eqs. (\ref{28}), (\ref{giango}) are in exact agreement with
$\langle U|T_{\mu\nu}|U\rangle$ given by eqs. (\ref{unin}), (\ref{unho})
once the luminosity $L$ is identified with
\be
\frac{L}{4\pi} =\frac{(\alpha^2-\beta^2)}{2(24M)^2}\ .
\label{lume}
\ee
\\ \noindent
iii) Israel-Hartle-Hawking state $|H\rangle$. \\ \noindent
This state is an equilibrium state regular both on the future and the
past horizons. For an equilibrium state $d=0$ which implies no net fluxes
($\langle H|S_{rt}(\phi)|H\rangle =0$). Inspection of eq. (\ref{25})
reveals that $\phi$ and $\partial_r\phi$
are regular both on the future
and the past horizons by imposing $E=0$ and $A=\alpha/24M$. This leaves the
parameter $B$ free. However, if the solution for $\phi$ is obtained by an
integral formula like eq. (\ref{conint}), which sets the lower limit of
integration to be $r=2M$, we would obtain
\be
 A=\frac{\alpha}{24M} \ ,\ \ \  B=\frac{11\alpha}{288M^2}\ ,\ \ \
 C=\frac{7\alpha M}{108} +\frac{\alpha M}{18}\ln 2M
\label{prpr}
\ee
 which implies $E=0$.
By this choice, which we think characterizes the $|H\rangle$ state, we
eventually arrive at the following expression
\begin{equation}
\label{lola}
\langle H|  S_{\mu}^{\ \nu}  |H \rangle \to
- \frac{k_2 \alpha}{96M^4}
\pmatrix{ 1 & 0 & 0 & 0
\cr 0 & 1 & 0 & 0
\cr 0 & 0 & 2 & 0
\cr 0 & 0 & 0 & 2
\cr}
\end{equation}
in the limit $r\to 2M$ and
\begin{equation}
\label{lopp}
\langle H|  S_{\mu}^{\ \nu} |H  \rangle \to
\frac{7}{2}\left(\frac{11}{864M^2}\right)^2(\alpha^2-\beta^2)
\pmatrix{ -1 & 0 & 0 & 0
\cr 0 & 1/3 & 0 & 0
\cr 0 & 0 & 1/3 & 0
\cr 0 & 0 & 0 & 1/3
\cr}
\end{equation}
as $r\to\infty$. The asymptotic limit of eq. (\ref{lopp}) describes
indeed a thermal bath as expected (compare to eq. (\ref{hhin})).
Note the appearance of the factor $(\alpha^2-\beta^2)$ just as in eq.
(\ref{giango}). \\
 Putting some numbers, if we choose $l_1=\frac{a}{8\pi\sqrt{-b}}$
which reconduces our $S_{eff}^{an}$ to Reigert's one, we have the
disappointing result $(\alpha^2-\beta^2)<0$ that is physically meaningless. A
similar situation was found in investigation of the anomaly induced effective
action representing 4D minimally coupled scalar fields spherically reduced to
2D at the classical level ( see \cite{treref}). For $l_1=0$ (i.e. complete
absorbtion of the conformally invariant part of our $S_{eff}^{an}$ in $S^W$)
we find that the overall coefficient of eq. (\ref{lopp}) is $\sim \frac{2.2\
10^{-4}}{\pi^2M^4} $ , which is much bigger than the correct one $\sim
\frac{8\ 10^{-6}}{\pi^2 M^4} $. Exact agreement of our result eq.
(\ref{lopp}) and eq. (\ref{hhin}) requires a fine tuning of the parameter
$\beta$, namely $\beta\sim\frac{6.2 \ 10^{-1}}{\pi} $. For this value of
$\beta $ the luminosity of an evaporating black hole (see eq. (\ref{lume}))
turns out to be $L= \frac{4.9\ 10^{-5}}{\pi M^2}$, which is roughly four
times smaller than the value given in \cite{dewitt}.
\\ \\ \noindent
Despite many nice features, there are some disappointing aspects of our
$\langle S_{\mu\nu}\rangle$. The overall coefficient of $\langle
B|T_{\mu\nu}|B\rangle$ on the horizon is expected to be the same
as the asymptotic limit of $\langle H|T_{\mu\nu}|H\rangle$ up to a minus
sign (see eqs. (\ref{boho}) and (\ref{hhin}) ). This is not the case for
$\langle S_{\mu\nu}\rangle$ (compare eq. (\ref{ciuco}) and eq.
(\ref{lopp}) ). \\
Unfortunately this is not all. If we go beyond the leading terms there is no
agreement of our $\langle S_{\mu\nu}\rangle$ and $\langle T_{\mu\nu}\rangle$.
Just to mention some negative aspects: $\langle U|S_{\theta}^{\
\theta}|U\rangle \sim r^{-3}$ in the limit $r\to \infty$ while  $\langle
U|T_{\theta}^{\ \theta}|U\rangle$ is expected to fall off as $r^{-4}$.
Furthemore in the same limit the coefficients in $r^{-1}$ and $r^{-2}$ of our
$\langle H|S_{\mu\nu}|H\rangle$ do not correspond to redshifted thermal
radiation as we have for $\langle H|T_{\mu\nu}|H\rangle$. One can guess that
this discrepancy may be removed by a careful modelling of the conformal part
in (\ref{coac}) through the introduction of other conformal structures
\cite{alter}, or, hopefully, by using the results of some of the direct
approximate calculations of the effective action (see, for example,
\cite{vilk}).

\sect{Conclusions}

In a field theory the knowledge of the trace anomaly allows part of the
effective action to be reconstructed; this is the so called anomaly induced
effective action, i.e. $S_{eff}^{an}$. The basic features of this object are
the nonlocality and the apparent absence of any reference to a particular
quantum state. Auxiliary fields have been introduced to make $S_{eff}^{an}$
local and boundary conditions have been imposed on them to select the
appropriate quantum states. \\ This procedure has been positively tested in a
simple 2D example where $S_{eff}^{an}$ is the well known Polyakov action. We
found exact agreement of the stress tensor calculated from this effective
action (i.e. $\langle S_{ab}\rangle$) and the one ($\langle T_{ab}\rangle$)
resulting from standard canonical quantization. \\ The expertise gained from
the analysis of this simple 2D model has allowed us to attack by similar
methods the physically much more interesting example of 4D Schwarzschild
black hole. We have been able to construct explicitly $\langle
S_{\mu\nu}\rangle$ for the three quantum states relevant for the discussion
of quantum matter fields in a black hole spacetime, namely Boulware state
(vacuum polarization around a static star), Unruh state (evaporation of a
black hole formed by gravitational collapse) and Israel-Hartle-Hawking state
(equilibrium of a black hole and a thermal bath). \\ By appropriate choice of
the arbitrary parameter $l_1$ of our model (or, equivalently, of $\beta$), we
were able to show the qualitative agreement of our $\langle S_{\mu\nu}\rangle
$ and the canonically computed $\langle T_{\mu\nu}\rangle$ at infinity and
near the horizon. Unfortunately these nice features cannot be extended beyond
leading terms. This is expected since the fondamental brick of our 4D
investigation, namely $S_{eff}^{an}$, up to a Weyl invariant term coincides
with Reigert's action. It is well known \cite{erosbo} that this action is
rather deficitary if considered as full effective action. Unlike Polyakov's
action Reigert's one is unable even to correctly reproduce the three-point
correlation function of the theory on the flat background. In view of this
fact, the discrepancies found in our $\langle S_{\mu\nu}\rangle$ are not
surprising at all. \\ What in our opinion is however remarkable is that,
neverthless, our $S_{eff}^{an}$ does indeed reproduce the expected behaviour
of the stress tensor near the horizon and at infinity upon which our
understanding of black hole evaporation is based. \\ In view of our
investigation, a careful use (not a straightforward one) of $\langle
S_{\mu\nu}\rangle$ in the semiclassical Einstein equations to get at least a
feeling of the backreaction in an evaporating black hole can be made. We hope
to come back to this point in a future work.

\vskip 4mm
\noindent
{\bf Acknowledgements} R.B. deeply thanks V. Frolov and R. Zucchini
for stimulating discussions.
A.F. is supported by an INFN fellowship.
I.Sh. is grateful to the Physics Department of UFJF for
warm hospitality, to CNPq for the fellowship and to FAPEMIG (MG)
for the travel grant.
His work was partially supported by RFFI (project 99-02-16617).


\begin{thebibliography}{99}

\bibitem{birdav}See for example N.D. Birrell and P.C.W. Davies,
{\it Quantum fields
in curved space}, Cambridge University Press, Cambridge (1982).

\bibitem{duff}
For a historical review see M.J. Duff, {\it Class. Quant. Grav.}
11 (1994), 1387.

\bibitem{Hawking}S. W. Hawking, {\it Nature} (London) 248 (1974), 30;
{\it Commun. Math. Phys.} 43 (1975), 199.

\bibitem{duff-77} S. Deser, M.J. Duff and C Isham,
             Nucl.Phys.{\bf 111B}, 45, (1976);
M.J. Duff, Nucl.Phys. {\bf 125B} 334 (1977).

\bibitem{chfu}S.M. Christensen and S.A. Fulling, {\it Phys. Rev.}
{\bf D15} (1977), 2088.

\bibitem{davies}P.C.W. Davies, {\it Proc. R. Soc. London} A354 (1977), 529.

\bibitem{dafu}P.C.W. Davies, S.A. Fulling and W.G. Unruh, {\it Phys. Rev.}
D13 (1976), 2720; P.C.W. Davies and S.A. Fulling, {\it Proc. R. Soc. London}
A354 (1977), 59.

\bibitem{molti}P. Candelas, {\it Phys. Rev. } D21 (1980), 2185;
D.N. Page, {\it Phys. Rev. } D25 (1982), 1499;
M.R. Brown, A.C. Ottewill and D.N. Page, {\it Phys. Rev. } D33 (1986), 2840;
V.P. Frolov and A.I. Zelnikov, {\it Phys. Rev.} D35 (1987), 3031;
C. Vaz, {\it Phys. Rev. } D39 (1989), 1776; {\it Phys. Rev.} D40 (1989),
1340;
P.R. Anderson, W.A. Hiscock and D.A. Samuel, {\it Phys. Rev.} D51 (1995), 4337.

\bibitem{bula}D.G. Boulware, {\it Phys. Rev. D11} (1975), 1404.

\bibitem{uuru}W.G. Unruh, {\it Phys. Rev.} D14 (1976), 870.

\bibitem{ishh}W. Israel, {\it Phys. Lett.} A57 (1976), 107;
J.B. Hartle and S.W. Hawking, {\it Phys. Rev. } D13 (1976), 2188.

\bibitem{full}S.A. Fulling, {\it J. Phys.} A10 (1977), 917.

\bibitem{dewitt}B.S. DeWitt {\it Phys. Rep.} C19 (1975), 297.

\bibitem{elster}T. Elster, {\it Phys. Lett.} A94 (1983), 205.

\bibitem{polyak}A.M. Polyakov, {\it Phys. Lett.} B103 (1981), 207.

\bibitem{book} I.L. Buchbinder, S.D. Odintsov and I.L. Shapiro,
 {\sl Effective Action in Quantum Gravity} (IOP, Bristol, 1992).

\bibitem{NEW}R. Balbinot and A. Fabbri, {\it 4D quantum black hole 
physics from 2D models?}, gr-qc/9904034.

\bibitem{reigert}R.J. Reigert, {\it Phys. Lett.} B134 (1984), 56.

\bibitem{frts}
E.S. Fradkin and A.A. Tseytlin, {\it Phys. Lett.} {\bf 134B}, 187 (1984).

\bibitem{antmot} I. Antoniadis and E. Mottola,
 Phys. Rev. {\bf 45D}, 2013 (1992).

\bibitem{cosh}
I.L. Shapiro and G. Cognola, Phys.Rev. {\bf 51D} 2775 (1995).

\bibitem{anju}
J.C. Fabris, A.M. Pelinson, I.L. Shapiro,
{\sl Anomaly-induced effective action for gravity and inflation}.
gr-qc/9810032.

\bibitem{pan}  S. Paneitz, A Quartic Conformally Covariant Differential
Operator for Arbitrary Pseudo-Riemannian Manifolds, MIT preprint, 1983.

\bibitem{shapiro} I.L. Shapiro and A.G. Jacksenaev, {\it Phys.Lett.}
{\bf 324B}, 284 (1994).

\bibitem{balfabsha}R. Balbinot, A. Fabbri and I. Shapiro, {\it Anomaly
induced effective actions and Hawking radiation}, hep-th/9904074.

\bibitem{treref} V. Mukhanov, A. Wipf and A. Zelnikov, {\it Phys. Lett.}
B332 (1994), 283; R. Balbinot and A. Fabbri, {\it Phys. Rev.}
D59 (1999), 044031;
F.C. Lombardo, F.D. Mazzitelli and J.G. Russo, {\it Phys. Rev.} D59 (1999),
064007.

\bibitem{erosbo}See for example J. Erdmenger and H. Osborn, {\it Nucl.
Phys.} B483 (1997), 431; S. Deser and A. Schwimmer, {\it Phys.Lett.}
{\bf B309} (1993) 279.

\bibitem{alter}J.A. de Barros and I.L. Shapiro
{\it Phys. Lett.} {\bf B412} (1997) 242.

\bibitem{vilk}  G.A. Vilkovisky, {\it Class.Quant.Grav.}
{\bf 9} (1992) 895;
A.O. Barvinsky, Yu. V. Gusev, G.A. Vilkovisky and Zhytnikov,
{\it Nucl.Phys.} {\bf B439} (1995) 561.
\end{thebibliography}
\end{document}